\documentclass[11pt,a4paper]{article}

\usepackage[a4paper,left=2cm,right=2cm,top=1cm,bottom=1.25cm,%
            includehead,headheight=50pt,headsep=10pt]{geometry}

\usepackage[T1]{fontenc}
\usepackage{helvet}   %
\usepackage{xcolor}

\usepackage{setspace}
\usepackage{graphicx}
\usepackage{caption}
\usepackage{float}
\usepackage{wrapfig}
\usepackage[hidelinks]{hyperref}

\usepackage{titlesec}
\titleformat{\section}{\normalfont\bfseries}{}{0pt}{}
\titlespacing*{\section}{0pt}{12pt}{8pt}
\titleformat{\subsection}{\normalfont\bfseries}{}{0pt}{}
\titlespacing*{\subsection}{0pt}{8pt}{3pt}
\usepackage{fancyhdr}
\newcommand{\conferencepageheader}{%
  \fancyhf{}%
  \renewcommand{\headrulewidth}{0pt}%
  \renewcommand{\footrulewidth}{0pt}%
  \fancyhead[L]{\raisebox{-0.3\height}{\includegraphics[width=6cm]{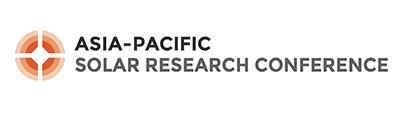}}}%
  \fancyhead[R]{\footnotesize 1--3 Dec 2026, Canberra}%
  \fancyfoot{}%
}
\fancypagestyle{conference}{\conferencepageheader}
\fancypagestyle{plain}{%
  \conferencepageheader
}
\begin{document}

\begin{center}
{\bfseries\large Distributed Multimodal Monitoring and Prediction for PV Asset and Site Conditions}

\end{center}

\begin{center}
Jan Stiefel, Brendan Do, Reza Arablouei, Ross Dungavell, Brano Kusy, Kenrick Anderson, \\ Benjamin Duck, Noel W Duffy, Peyman Moghadam

CSIRO

\textit{firstname.lastname@csiro.au}
\end{center}

\vspace{3pt}

\section{Abstract}

Utility-scale photovoltaic (PV) monitoring remains fragmented: electrical performance, structural condition and vegetation management are typically assessed through separate workflows, making it difficult to relate changes in energy performance to the local physical and environmental conditions experienced by individual modules. We propose a distributed multimodal monitoring and prediction framework that combines mechanical, optical, electrical, and microclimate measurements at the string level. Rather than treating these measurements as independent sensor streams, the framework uses them to derive higher-level indicators of asset and site condition, including quantities that are not directly instrumented. We designed and installed four compact wireless nodes attached directly to the rear of PV modules for a three-month pilot study to demonstrate module-level monitoring. From low-cost multimodal measurements, the system detected vegetation-management events, predicted wind speed from module vibration on unseen data, and extracted repeatable structural-frequency fingerprints associated with individual module locations. These preliminary results demonstrate a pathway from distributed monitoring toward predictive PV asset management, where multimodal observations can be used to predict or estimate otherwise unobserved operating and structural states, identify emerging changes earlier, and provide more granular intelligence than plant-level electrical monitoring or periodic inspection alone.

\section{Problem Statement}

Utility-scale PV farms are large, spatially distributed assets in which relatively small module- or string-level degradation can accumulate into meaningful losses in energy yield and revenue. Existing plant monitoring is predominantly based on electrical measurements aggregated at inverter or combiner level. These measurements are effective for identifying production loss, but often provide limited information about where the underlying problem is located, what physical process caused it, or whether a condition is developing before production is affected. Structural condition, vegetation, and local environmental effects are typically assessed through separate inspection workflows~\cite{Silitonga2024}, making it difficult to connect plant-level performance changes with module-level physical causes~\cite{Tang2024}. This creates a need for low-cost, persistent, spatially granular monitoring that can infer otherwise unobserved asset and site conditions and support earlier, predictive maintenance decisions~\cite{Jahn2022}.

\section{Pilot Deployment}

\begin{wrapfigure}[10]{r}{0.4\textwidth}
  \vspace{-32pt}
  \centering
  \includegraphics[width=0.32\textwidth]{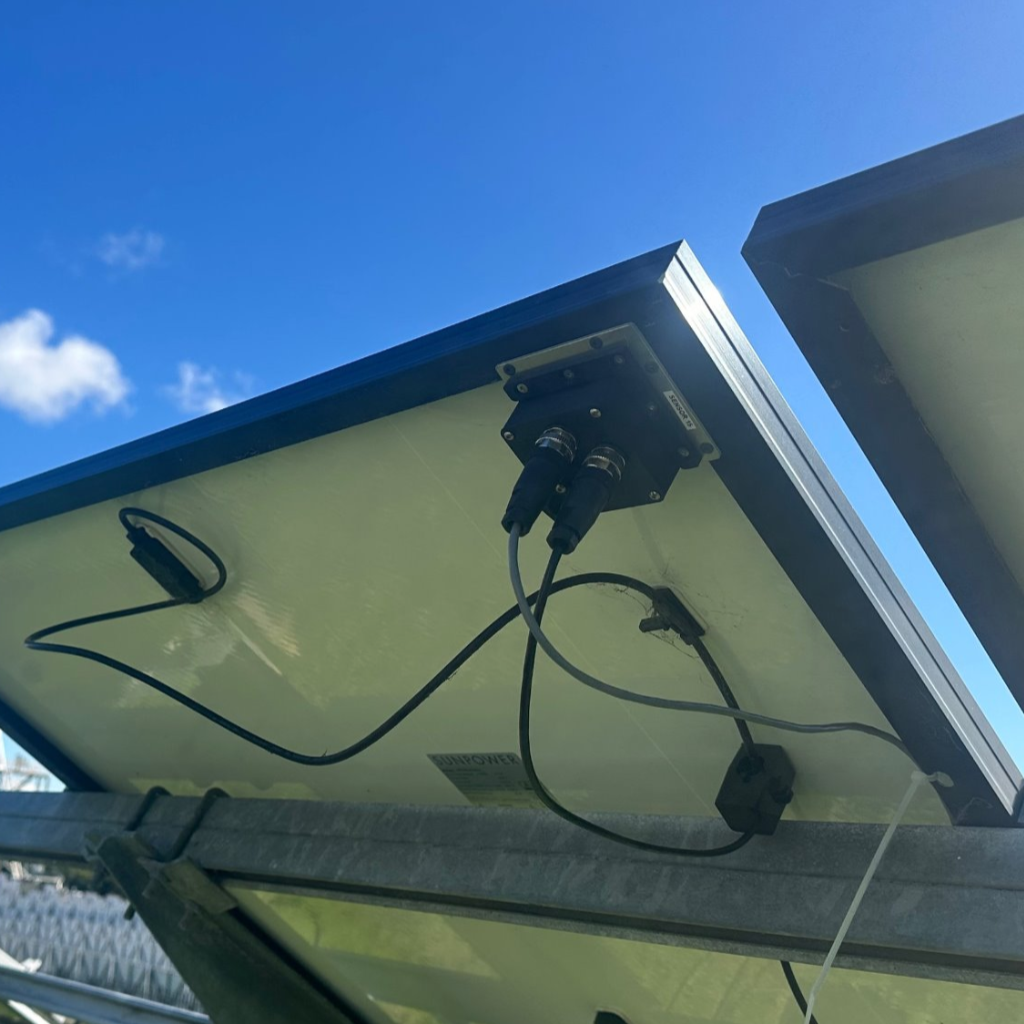}
  \caption{Compact wireless node mounted on the rear of a PV module.}
  \label{fig:device}
  \vspace{-8pt}
\end{wrapfigure}

We designed compact, wireless, multimodal nodes for persistent monitoring at the individual PV module level. Each node integrates mechanical, optical, electrical, and microclimate measurements, including a 100 Hz triaxial accelerometer, downward-looking colour/infrared sensing, temperature, humidity, pressure, and non-invasive DC current sensing. The node supports local storage and long-range, low-bandwidth wireless communication.

Four prototype nodes were then attached directly to the rear of PV modules at CSIRO's Photovoltaic Outdoor Research Facility in Newcastle, Australia. Two nodes were mounted at different positions on the same module, while two additional nodes were installed at comparable positions on adjacent modules. This arrangement enabled comparison across both mounting locations and nominally similar modules while collecting continuous multimodal observations over approximately three months.

The accelerometer is rigidly coupled to the module, allowing its quasi-static response to provide module attitude estimates while higher-frequency components capture structural vibration. The downward-looking optical sensor observes under-array vegetation, and a co-located weather station provides an independent reference for wind conditions. Unless otherwise stated, results are reported over a 3-month analysis window, June-August 2026, for which data were available across all four nodes and the main sensing modalities. %

\section{Preliminary Results}

The four nodes achieved a mean packet reception ratio of 99.98\% across all nine sensing channels, providing a near-continuous dataset for cross-node and multimodal comparison. Because all channels are transmitted together in each device record, the consistent reception rates also indicate that no individual sensing channel experienced a separate data-loss issue.

The deployment demonstrates three examples of higher-level condition prediction from module-level multimodal observations: under-array vegetation condition, wind-speed prediction from structural vibration, and structural-condition fingerprinting. Together, these results show how low-cost distributed measurements can be transformed into indicators of asset and site states that are otherwise not directly instrumented.

\subsection{Under-array vegetation condition monitoring}

\begin{figure}%
  \centering
  \includegraphics[width=\textwidth]{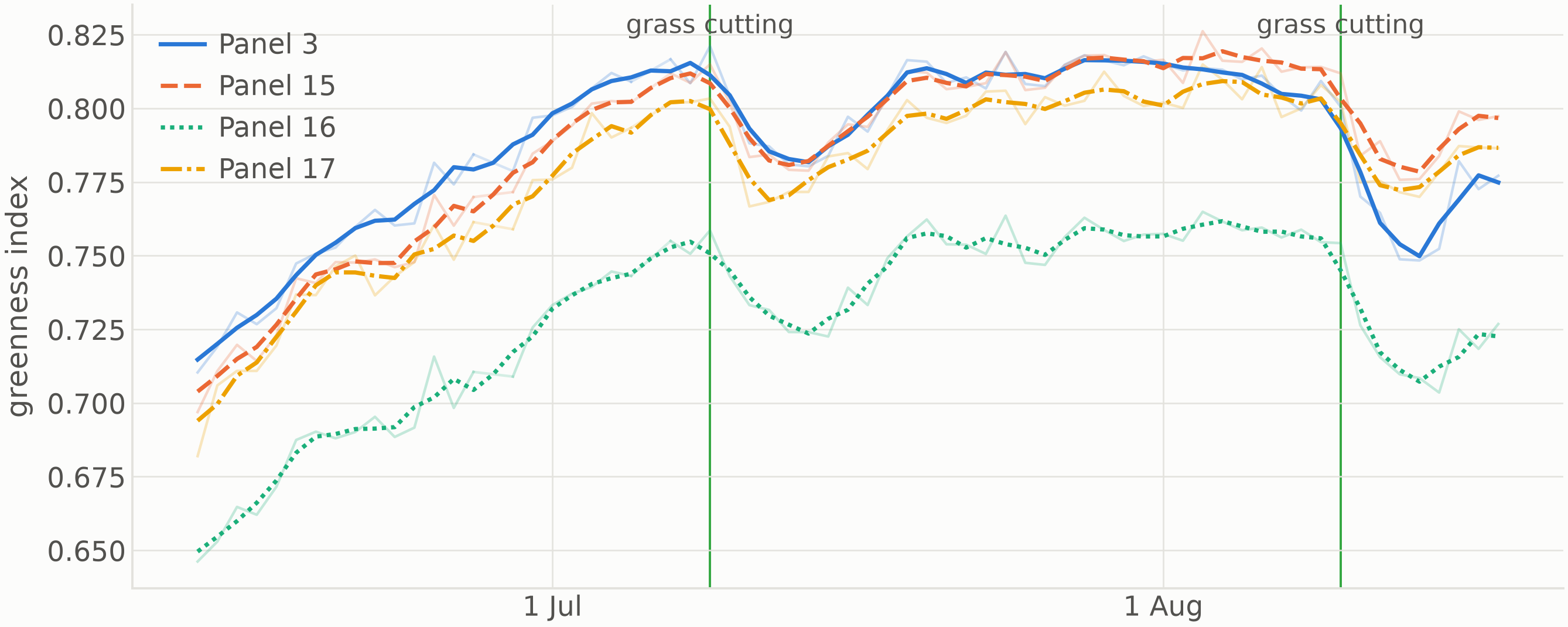}
  \caption{Predicted under-array vegetation condition using an optical greenness index as a proxy for relative biomass beneath four monitored PV modules.} %
  \label{fig:greenness}
\end{figure}

Figure~\ref{fig:greenness} shows the predicted under-array vegetation condition derived from an infrared-to-red greenness index. The four traces share a seasonal rise and two distinct declines, both aligned with recorded vegetation-management events. This demonstrates that a simple non-imaging optical measurement can provide a continuous proxy for relative vegetation biomass beneath individual PV modules. Differences in absolute values, particularly for node 16, are influenced by mounting position, illumination, and viewing geometry; therefore, the relative change of the index is more informative than its absolute magnitude.

\subsection{Wind speed prediction from module vibration}

\begin{figure}%
  \centering
  \includegraphics[width=\textwidth]{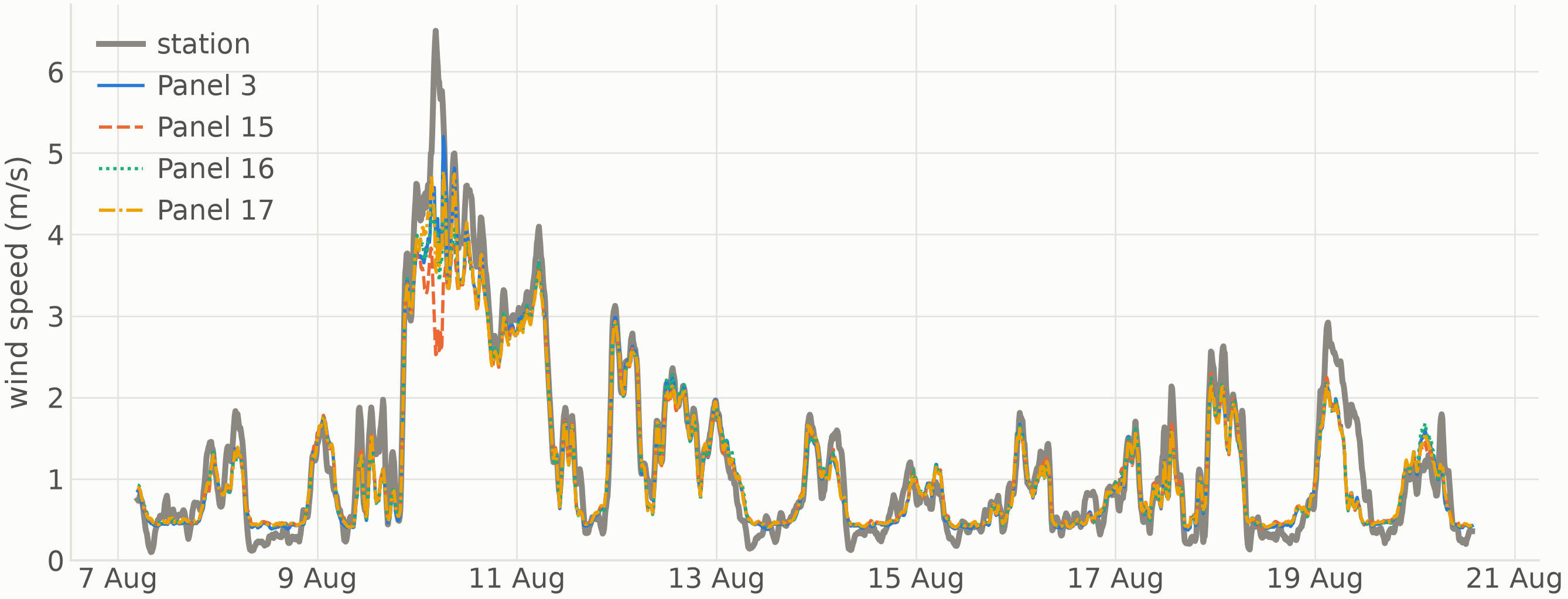}
  \caption{Vibration-based wind-speed predictions from PV modules compared with co-located weather-station ground truth over the chronological hold-out period.}%
  \label{fig:wind}
\end{figure}

Figure~\ref{fig:wind} compares wind-speed predictions derived from module vibration against the co-located weather station used as ground truth. On the chronological hold-out period, the vibration-based model achieved $R^2=0.738$ and closely tracked repeated moderate-wind events. The largest ground-truth wind-speed peak reached approximately $6.5\,m/s$, while predictions reached around $4.5\,m/s$, indicating underestimation at higher wind speeds.
The temporal correspondence demonstrates that module vibration contains sufficient information to predict local wind conditions. This provides a pathway to estimating spatially distributed environmental loading across a utility-scale PV farm without requiring dedicated wind instrumentation at each module.

\subsection{Structural frequency fingerprints}

\begin{figure}%
  \centering
  \includegraphics[width=\textwidth]{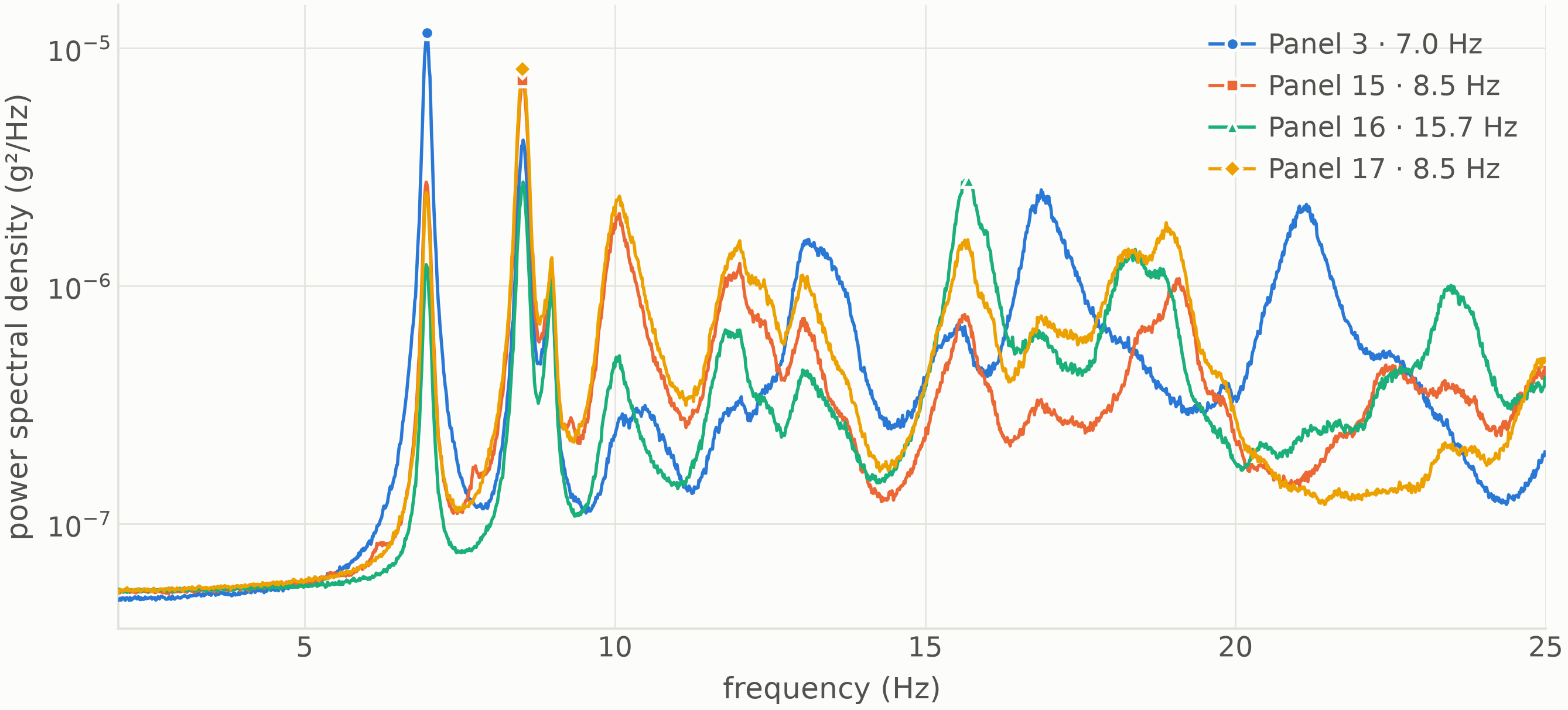}
  \caption{Median vibration spectra during the windiest periods. Recurring resonances and differences in relative amplitude form a candidate structural signature for each location.}
  \label{fig:fingerprint}
\end{figure}

Figure~\ref{fig:fingerprint} shows repeatable vibration spectra measured across the monitored PV modules during high-wind periods. Several resonance peaks are shared across nodes due to the common rack structure, while differences in relative amplitude reflect local mounting position and structural response. Nodes 15 and 17, mounted at comparable positions on adjacent identical modules, exhibit similar spectral fingerprints, including a prominent peak near $8.5\,Hz$.

These repeatable frequency-domain patterns provide a baseline representation of local structural condition. Persistent changes in peak frequency, the appearance or disappearance of resonances, or sustained changes in relative amplitude could indicate changes in mounting stiffness, clamp condition, or structural integrity. Controlled fault experiments are required to determine how reliably these spectral changes can predict specific structural failure modes.

\section{Conclusion \& Future Work}

This work demonstrates that low-cost, module-mounted multimodal nodes can provide more than raw measurements: they can support prediction of otherwise unobserved asset and site conditions at fine spatial and temporal resolution. Across a three-month deployment, the same distributed platform captured vegetation-condition changes, predicted local wind speed from structural vibration, and produced repeatable structural fingerprints at individual PV modules. These results establish an initial pathway from fragmented monitoring toward persistent, module-level intelligence for PV asset management.  Validation is limited to four nodes at one site over three months; performance across seasons, sites and mounting configurations remains to be established.
The broader opportunity is not to replace inverter monitoring, ground-based or aerial robotic inspection, but to complement them with persistent, spatially distributed physical context at module level. A common multimodal platform could provide local indicators of asset conditions, environmental loading, structural change, vegetation activity and electrical behaviour, helping connect plant-level performance anomalies to local physical causes while retaining low-bandwidth scalability. Future work will focus on turning these indicators into predictive-maintenance capabilities through longer deployments across seasons, sites and rack configurations, controlled perturbations and maintenance events for ground-truth validation, and optimisation of node placement and density to balance spatial coverage, redundancy and prediction accuracy. Multimodal fusion and on-node or gateway inference will further enable compact health indicators and anomaly events to be communicated instead of continuous raw data.

\end{document}